\documentclass{an}
\usepackage{graphicx}
\usepackage{times}
\usepackage{fancyhdr}
\usepackage{supertabular}

\begin{document}

\title{Follow-up observations of Comet 17P/Holmes after its extreme outburst in brightness end of October 2007\thanks{Based on observations obtained with telescopes of the University Observatory Jena, which
is operated by the Astrophysical Institute of the Friedrich-Schiller-University.}}

\author{Markus~Mugrauer\inst{1}, M.~M.~Hohle\inst{1}$^{,}$\inst{2}, C.~Ginski\inst{1}, M.~Vanko\inst{1}, F.~Freistetter\inst{1}}

\institute{Astrophysikalisches Institut und Universit\"ats-Sternwarte Jena, Schillerg\"asschen 2-3,
07745 Jena, Germany \and Max-Planck-Institut f\"{u}r extraterrestrische Physik, Giessenbachstrasse,
85748 Garching, Germany}

\date{Received; accepted; published online}

\abstract{We present follow-up observations of comet 17/P Holmes after its extreme outburst in
brightness, which occurred end of October 2007. We obtained 58 V-band images of the comet between
October 2007 and February 2008, using the Cassegrain-Teleskop-Kamera (CTK) at the University
Observatory Jena. We present precise astrometry of the comet, which yields its most recent
Keplerian orbital elements. Furthermore, we show that the comet's coma expands quite linearly with
a velocity of about 1650\,km/s between October and December 2007. The photometric monitoring of
comet 17/P Holmes shows that its photometric activity level decreased by about 5.9\,mag within 105
days after its outburst. }
\correspondence{markus@astro.uni-jena.de}

\maketitle

\section{Introduction}

Comet 17P/Holmes is a short periodic comet, which was discovered by the British amateur astronomer
Edwin Holmes close to the bright Andromeda galaxy M31 on November 6th 1892 (see \cite{holmes1892}
or \cite{krueger1893}). The comet passed through its perihelion already several month before its
discovery, and had been in a much better condition for discovery before. Nevertheless, it was not
detected which points out that the comet experienced most probably a strong outburst in brightness
at the end of November 1892, which eventually enabled its visual discovery (\cite{boss1893}).
\cite{barnard1896}, who carried out a follow-up study of the comet after its discovery, described
that on November 9th 1892 17/P Holmes appeared to the naked eye brighter than the central part of
M31, with a coma diameter of about 6\,arcmin. During the next weeks the comet faded rapidly.
However, mid of January 1893 a further outburst occurred and the magnitude of 17/P Holmes increased
again to the 8th magnitude (\cite{campbell1893}). Afterwards, the comet steadily became fainter,
and no further successful observations were reported later than begin of April 1893.

As member of the Jupiter family, close encounters between the gaseous giant planet Jupiter and 17/P
Holmes take place occasionally. These encounters cause changes of the comet's orbit and have to be
taken into account in the calculation of the comet's ephemerides. The first close encounter with
Jupiter in the 20th century occurred 1908 (\cite{kronk2003}). 17P/Holmes was observed successfully
during its perihelion passages in 1892, 1899 and 1906. However, after its encounter with Jupiter,
in 1908, the comet got lost and could not be detected during its expected following perihelion
passages until that of 1964. Only by taking into account all astrometric measurements of the comet
before it got lost after 1906, as well as the gravitational perturbations of Jupiter, it was
possible to determine the altered orbital elements, and therefore to calculate more accurate
ephemerides of the comet. Eventually, 17P/Holmes was recovered in 1964 and could be imaged during
all its following perihelion passages.

The last time 17P/Holmes passed through its perihelion on May 4th 2007 (see section 3). The comet's
brightness ranged between 14.1 and 16.5\,mag between mid of July and begin of October 2007. On
October 24th J. A. Henriquez Santana observed the comet and reported that it appeared much brighter
($m\sim8.4$\,mag) than just a few days before (\cite{buzzi2007}). The brightness of the comet
continued to increase and finally reached about 2.5\,mag, when 17/P Holmes appeared on the night
sky as a yellow shiny point like object, easily visible to the naked eye. Hence, the brightness of
the comet increased from about 16.5\,mag up to 2.5\,mag, i.e. in total an outburst in brightness of
14\,mag, the largest outburst of a comet observed so far.

Such a huge outburst in brightness could be caused by a meteorite impact on the comet's surface,
which is however less probable, in particular if one takes into account that the comet showed
already two outbursts before, end of 1892, and in early 1893. Another, more favorable scenario, is
that, the comet was heated up by solar radiation during its perihelion passage, and underground
layers were evaporated. Later on, the produced internal gas pressure disrupted the comet surface,
which led to a dramatic emission of gas and dust, causing the outburst of the comet.

In this paper we present results of our follow-up observations of 17P/Holmes after its outburst. In
section 2 we describe in detail all observations, which were carried out. In the following section
"Data Analysis and Discussion" we present accurate astrometry of the comet, which allows the
determination of its most recent orbital elements. Furthermore, we report on the evolution of the
comet's coma, as well as on its photometric activity level after the outburst.

\section{Observations}

\begin{figure}[h]
\resizebox{\hsize}{!}{\includegraphics[]{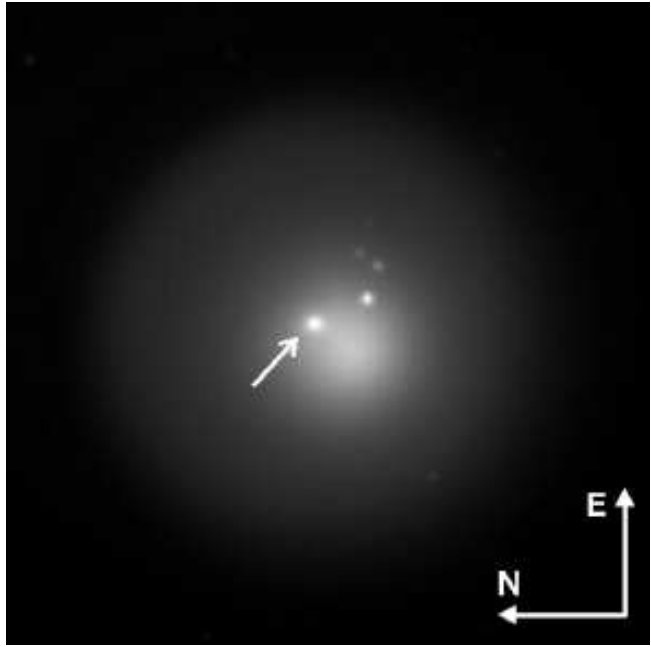}} \caption{Comet 17/P Holmes on October
29th 2007. The total integration time of this CTK image is 80\,s in V-band. In order to avoid
saturation of the CTK detector on the bright comet, four CTK images, each with 20\,s of integration
time, are taken, aligned, and finally added up. The pattern shown here is a display of the whole
CTK image with 8.3\,$'$$\times$8.3\,$'$ field of view. The bright central part of the comet's coma
is indicated with a white arrow. Several faint stars are visible, which shine through the outer
thinner part of the coma, whose apparent angular radius measures already $\sim 3.6$\,arcmin.}
\label{holmes291007}
\end{figure}

After its dramatic outburst in brightness end of October 2007, we started an imaging campaign of
comet 17/P Holmes at the University Observatory Jena, using the Cassegrain-Telescope-Kamera
(CTK)\footnote{See \cite{mugrauer2009} for a detailed description of this CCD imager.}. The comet
could be observed with the CTK the first time on October 29th 2007, i.e. only five days after the
discovery of the outburst. In total 58 images of the comet are taken, spread over 8 nights between
October 2007 and February 2008. All observations are carried out in V-band and different
integration times are used to avoid saturation of the CTK detector on the bright comet. In our
first observing epoch integration times of only 20, and 30\,s are applied, due to the high
brightness of the comet (see Fig.\,\ref{holmes291007}). Because of the fast decrease of the comet's
brightness longer integration times of 60, and 300\,s are chosen in the following observing epochs
(see Fig.\,\ref{holmes011207}).

In order to remove the bias level, as well as the dark current from all our CTK images, at least
three darkframes are always taken directly after the imaging of the comet. Cosmics are effectively
suppressed in the median of these darkframes, and therefore the median of the darkframes is always
subtracted from our CTK images. Afterwards, all images are flatfielded with skyflats, which are
always taken in the same nights as our comet observations, either in evening or morning twilight.
Furthermore, to remove bad pixels, we apply a badpixel correction to all dark+bias subtracted and
flatfielded CTK images. We use a badpixel mask, which is created out of a series of domeflats,
taken with different exposure times. The whole data-reduction is done with \textsl{ESO-MIDAS} and
standard \textsl{IRAF} scripts.

All CTK images of comet 17/P Holmes are astrometrically calibrated using reference sources which
are detected together with the comet in our CTK images, and are listed in the 2MASS Point Source
Catalogue (\cite{skrutskie2006}). After calibration, the averaged uncertainty of the achieved
absolute astrometry of our CTK images is only 0.28\,arcsec relative to the 2MASS Point Source
Catalogue. The astrometric calibration yields a pixelscale of the CTK detector of
2.2064$\pm$0.0003\,arcsec/pixel for all observing epochs, i.e. the CTK field of view measures
37.7\,$'$$\times$\,37.7$'$.

\section{Data Analysis and Discussion}

In order to derive the most recent Keplerian elements of the orbit of comet 17/P Holmes, we measure
the comet's position in all our 58 CTK images, using the gaussian centering algorithm
"center/gauss" of \textsl{ESO-MIDAS}. The derived astrometry of the comet for all CTK observations
is summarized in Tab.\,\ref{tabastro}. We fit a Keplerian orbit to the measured comet's positions,
and obtain its most recent orbital elements, which are summarized in Tab.\,\ref{tabelem}. The RMS
error of the obtained orbital fit is only 0.663\,arcsec. According to our astrometric measurements
comet 17/P Holmes is on an eccentric ($e=0.43277\pm0.00003$) orbit, and revolves around the Sun
with a period of $P=6.8852\pm0.0004$\,yr. The comet's last perihelion passage took place on May 4th
2007 ($JD_{Perihel}=2454225.06\pm0.02$) when it was separated from the Sun only by
$q=2.0529\pm0.0001$\,AU. The next perihelion passage of the comet is predicted for March 21th 2014
($JD_{Perihel}=2456738$).

\begin{table}[h!]
\centering\caption{Orbital elements of comet 17P/Holmes, derived by fitting an Keplerian orbit to
our astrometry of the comet. The RMS error of the obtained orbital fit is 0.663\,arcsec.}
\label{tabelem}
\begin{tabular}{ll}\hline
$JD_{Perihel}$      & 2454225.06$\pm$0.02\\
$a$                 & 3.6192$\pm$0.0001\,AU\\
$e$                 & 0.43277$\pm$0.00003\\
$i$                 & 19.1119$\pm$0.0001\,$^{\circ}$\\
$\omega$            & 24.266$\pm$0.002\,$^{\circ}$\\
$\Omega$            & 326.8666$\pm$0.0009\,$^{\circ}$\\
\hline\hline
\end{tabular}
\end{table}

With the derived pixelscale of the CTK detector we can determine the extent of the coma of comet
17/P Holmes in all our CTK images. However, the whole extent of the comet's coma is imaged only in
the first observing epoch in October 2007. Due to its fast expansion the size of the coma already
exceeds the CTK field of view in all following observing epochs. In order to image completely also
the widely expanded coma of comet 17/P Holmes, we obtained a mosaic of 9 CTK images on December
1$^{\rm th}$ 2007. This mosaic, which covers a total field of view of
1.57\,$^{\circ}$$\times$1.57\,$^{\circ}$, is shown in Fig.\,\ref{holmes011207}. The total angular
radius $\rho$ of the comet's coma, measured in both CTK observing runs in October and December
2007, is summarized in Tab.\,\ref{tabcoma}. The same table also lists the comet's Earth distance
$\Delta$, calculated with the derived orbital elements (see Tab.\,\ref{tabelem}).

\begin{table}[h!]
\centering\caption{Expansion of the coma of comet 17P/Holmes. This table lists the total ($\rho$)
and the minimum angular radii ($\rho_{min}$) of the comet's coma for all observing epochs at which
the radii can be measured in the CTK images. With the comet's Earth distance $\Delta$, derived from
the determined orbital elements of the comet, the projected total radius $R$ of the coma, and its
projected minimum radius R$_{min}$ can be derived.} \label{tabcoma}
\begin{tabular}{lccc}\hline
epoch & $\rho$         & $\Delta$ & $R$\\

      & [arcsec]       & [AU]     & [10$^{\rm{6}}$km]\\
\hline
29/10/07 & 216$\pm$7   & 1.6265 & 0.256$\pm$0.009\\
01/12/07 & 1293$\pm12$ & 1.6967 & 1.595$\pm$0.015\\
\hline\hline
epoch & $\rho_{min}$ & $\Delta$ & $R_{min}$\\

      & [arcsec]       & [AU]     & [10$^{\rm{6}}$km]\\
\hline
29/10/07 & 185$\pm$7  & 1.6265 & 0.219$\pm$0.009\\
27/11/07 & 872$\pm$9  & 1.6741 & 1.061$\pm$0.011\\
28/11/07 & 891$\pm$7  & 1.6792 & 1.089$\pm$0.009\\
01/12/07 & 953$\pm$12 & 1.6967 & 1.176$\pm$0.015\\
\hline\hline
\end{tabular}
\end{table}

Furthermore, Tab.\,\ref{tabcoma} shows the projected radius $R$ of the coma, derived with its
angular radius, and the comet's Earth distance for each observing epoch. With the given epoch
difference between our two CTK observations in October and December 2007 ($\Delta t=33.81$ days),
we can finally determine the expansion velocity of the comet's coma and obtain
$v_{exp}=1651\pm22$\,km/h.

\begin{figure}[h!]
\resizebox{\hsize}{!}{\includegraphics[]{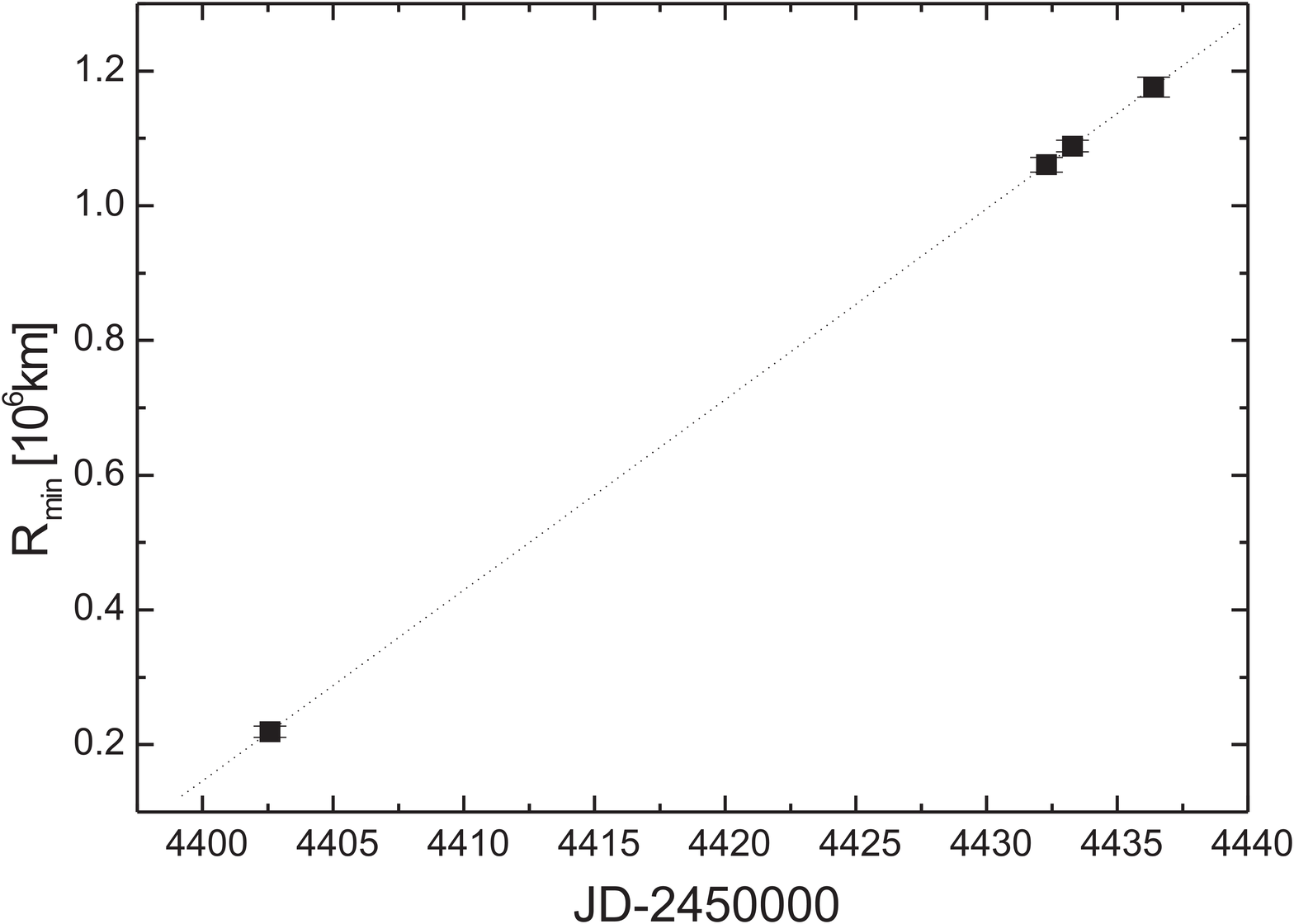}} \caption{The expansion of the coma of
comet 17/P Holmes during the first month after its outburst in brightness. This plot shows the
evolution of the minimum radius $R_{\rm{min}}$ of the comet's coma, between end of October until
begin of December 2007. The minimum radius expands linearly over time with a velocity of
1179.6$\pm$0.5\,km/s, the slope of this linear fit, which is illustrated as dotted line.}
\label{comaexpand}
\end{figure}

Although, the whole coma is not imaged in all observing epochs, we can measure its minimum angular
radius $\rho_{min}$ in the first four epochs. The derived projected minimum radius of the comet's
coma is listed in the second part of Tab.\,\ref{tabcoma} and is plotted for a range of time in
Fig.\,\ref{comaexpand}. The minimum radius of the coma expands quite linearly over time. A linear
fit, which is illustrated with a dotted line in Fig.\,\ref{comaexpand}, yields a slope of
1179.6$\pm$0.5\,km/h. This increase of $R_{min}$ can be considered as the minimum expansion
velocity of the comet's coma. Due to the solar wind the coma of the comet should not expand with
the same velocity in all directions. In particular, in direction towards the Sun we expect a
minimum expansion velocity, resulting in a non-radial symmetric shape of the coma. If we assume
that the coma expands linearly over time the outburst time can be derived directly from the linear
fit of the expansion of the coma. According to the expansion of $R_{min}$ we obtain an outburst
time $JD_{out}=2454394.8\pm2.3$, i.e. October $21.3 \pm 2.3$. If we use the expansion of the total
radius of the coma we obtain an outburst time $JD_{out}=2454396\pm1$, i.e. October $22.5 \pm 1$.
Hence, we can conclude that the outburst started only few days or even only few hours before it was
discovered on October 24th.

\begin{figure}[h!]
\resizebox{\hsize}{!}{\includegraphics[]{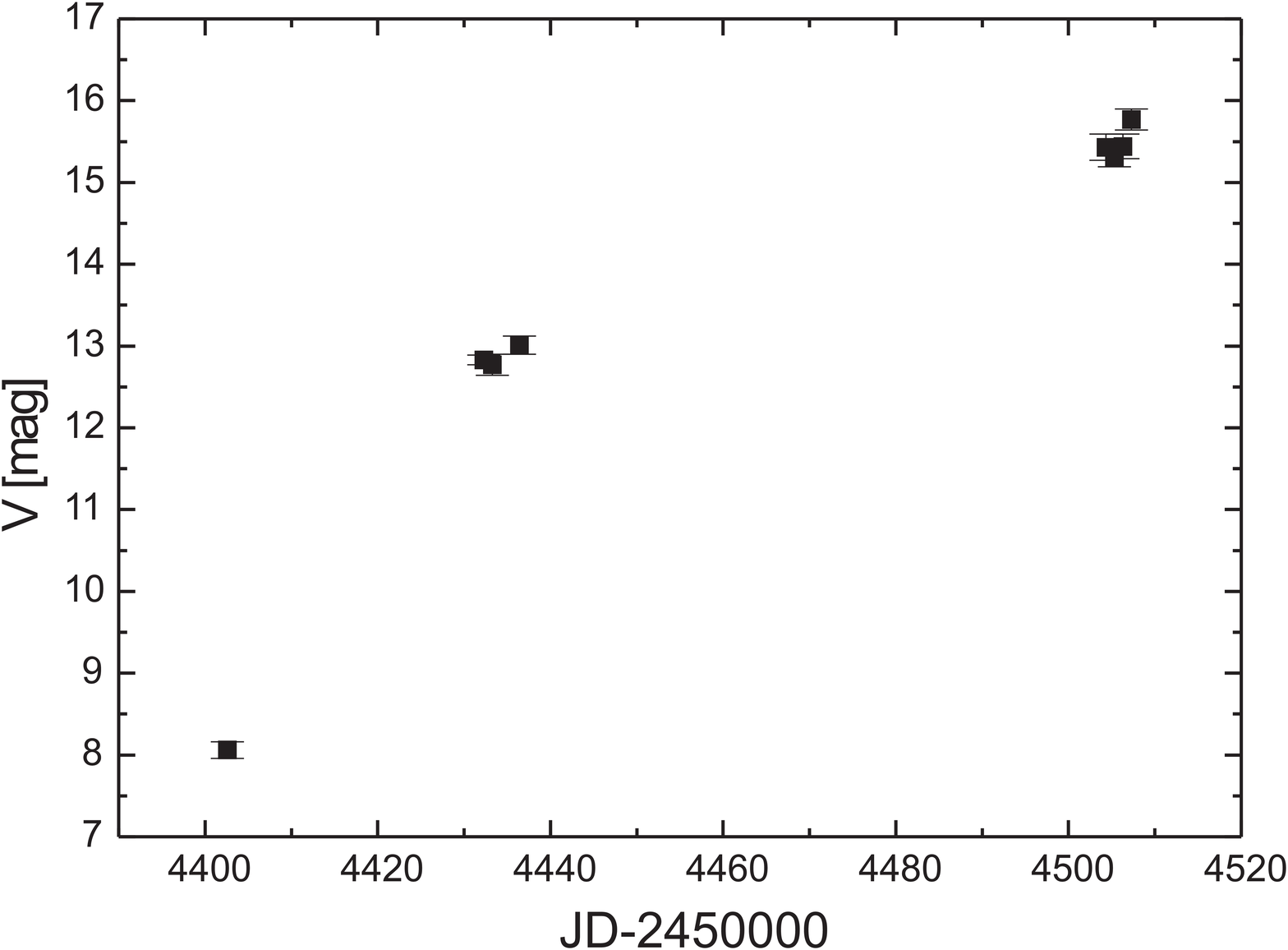}} \caption{This plot shows the apparent V-band
magnitude $V$ of the central coma of comet 17P/Holmes for all CTK observing epochs between October
2007 and February 2008. Between 105 days its apparent brightness decreases by $\Delta
V$=7.71$\pm$0.16\,mag.} \label{photo}
\end{figure}

Beside the astrometric measurements described above, we also monitored the evolution of the comet's
photometry after its outburst in brightness. In Tab.\,\ref{tabphoto1} we list the measured apparent
V-band magnitudes of the bright central part of the comet's coma, which is also plotted for a range
of time in Fig.\,\ref{photo}. For photometric calibration of the individual CTK images we always
use several reference stars from the Hipparcos and Tycho catalogues (\cite{perryman1997}), which
are imaged together with the comet and exhibit an accurate V-band photometry. With this method a
photometric accuracy of $\Delta m=0.12$\,mag can be achieved, in average. The apparent V-band
brightness of the comet's central coma decreases by about 7.7\,mag during 105 days between October
29th 2007 and February 10th 2008. Because also the comet's Earth distance $\Delta$ increases
significantly during this period (see Tab.\,\ref{tabphoto1}), the change of distance has to be
taken into account to obtain the true photometric activity level of the comet. Therefore, we
determine the absolute magnitude $M_{\rm V}$ of the comet's central coma, i.e. its apparent
magnitude at a fixed distance of 1\,AU. The absolute magnitude can be derived with the measured
apparent magnitude of the coma and the comet's Earth distance $\Delta$. Tab.\,\ref{tabphoto1} shows
the derived absolute magnitude of the comet's central coma for all observing epochs. The accuracy
of the derived absolute photometry is dominated by the accuracy of the measured apparent
magnitudes.

\begin{table}[h!]
\centering\caption{Photometry of the central coma of comet 17/P Holmes for all CTK observing
epochs. The apparent V-band magnitude of the coma is listed together with its absolute magnitude,
derived with its apparent photometry and the calculated comet's Earth distance $\Delta$.}
\label{tabphoto1}
\begin{tabular}{lccccc}\hline
epoch & JD-2450000 & $V$ & $\Delta$ & M$_{\rm V}$\\
& & [mag] & [AU] & [mag]\\
\hline
29/10/07 & 4402.58 & \,\,\,8.06$\pm$0.10 & 1.6265 &  \,\,\,7.00$\pm$0.10 \\
27/11/07 & 4432.30 & 12.83$\pm$0.06 & 1.6741 & 11.71$\pm$0.06 \\
28/11/07 & 4433.29 & 12.77$\pm$0.13 & 1.6792 & 11.65$\pm$0.13 \\
01/12/07 & 4436.39 & 13.01$\pm$0.11 & 1.6967 & 11.86$\pm$0.11 \\
07/02/08 & 4504.31 & 15.43$\pm$0.16 & 2.5433 & 13.40$\pm$0.16 \\
08/02/08 & 4505.29 & 15.30$\pm$0.11 & 2.5597 & 13.26$\pm$0.11 \\
09/02/08 & 4506.29 & 15.44$\pm$0.15 & 2.5764 & 13.39$\pm$0.15 \\
10/02/08 & 4507.27 & 15.77$\pm$0.13 & 2.5930 & 13.71$\pm$0.13 \\
\hline\hline
\end{tabular}
\end{table}

During our follow-up observations the absolute magnitude of the central coma of comet 17/P Holems
decreased by $\Delta M_{V}=6.71\pm0.16$\,mag, see Tab.\,\ref{tabphoto1}. However, because of the
increase of the comet's Sun distance in the same span of time, the comet's natural activity level
is also expected to decrease, due to lower solar radiation at larger solar distance.

\cite{whipple1978} studied the natural brightness variation of comets before and after their
perihelion passage which follows in general a $r^{-n}$ relation ($n>2$)\footnote{The expected
natural decrease of brightness of a non-active body should follow a $r^{-2}$ relation, which is
also expected for comets but only at very large solar distances $r$.}. For short periodical comets,
essential for those of the Jupiter family, he reported comparable $n$-values before and after their
perihelion passages, namely $5.0\pm1.4$, and $4.61\pm0.91$, respectively. Taking into account the
decrease of the dust production rate $Q_{D}$ at wider Sun distances these empirical relations agree
well with the theoretical approach derived by \cite{fernandez1999}, which yielded a $Q_{D}^{2}
\cdot r^{-4}$ dependency.

As comet 17/P Holmes already passed through its perihelion in May 2007, i.e. before our CTK
observations, we use the $r^{4.61\pm0.91}$ relation of \cite{whipple1978} to approximate the
expected natural decrease of the absolute brigntness of the comet's central coma. The difference
between the decrease of the absolute magnitude of the coma and its expected natural decrease,
finally yields the true decrease of the comet's outburst activity level. Within the 105 days of our
follow-up observations of comet 17/P Holmes, its distance to the Sun increased from 2.46 to
2.89\,AU. Therefore, according to the relation reported by \cite{whipple1978}, we expect a natural
decrease of the absolute magnitude of the comet's central coma of only $\Delta
M_{V\star}=0.82\pm0.16$\,mag, while the observed decrease is $\Delta M_{V}=6.71\pm0.16$\,mag.
Hence, the outburst activity level of 17/P Holmes significantly decreases by $\Delta
M=5.89\pm0.23$\,mag, i.e. a factor between 183 to 281 within the given span of time.
Tab.\,\ref{tabphoto2} shows the observed decrease of the absolute magnitude of the comet's central
coma, its expected natural magnitude decrease, as well as the derived decrease of the comet's
outburst activity level, for all CTK observing epochs.

\begin{table}[h!]
\centering\caption{Photometric variability of the central coma of comet 17/P Holmes since its
outburst in brightness. The comet's Sun distance $r$ is calculate with the derived orbital elements
of the comet. $\Delta M_{V}$ is the observed decrease of the absolute magnitude of the comet's
central coma since the first CTK observing epoch. $\Delta M_{V\star}$ is the expected natural
decrease of its absolute magnitude due to the change of the comet's Sun distance. Their difference,
$\Delta M$, yields the decrease of the comet's outburst activity level.} \label{tabphoto2}
\begin{tabular}{lcccc}\hline
epoch & $r$     & $\Delta M_{V}$ & $\Delta M_{V\star}$ & $\Delta M$\\
      & [AU]    & [mag]              & [mag]           & [mag]\\
\hline
29/10/07 & 2.4553 & 0                & 0             & 0\\
27/11/07 & 2.5738 & 4.71$\pm$0.12    & 0.24$\pm$0.05 & 4.47$\pm$0.13\\
28/11/07 & 2.5778 & 4.65$\pm$0.16    & 0.24$\pm$0.05 & 4.41$\pm$0.17\\
01/12/07 & 2.5906 & 4.86$\pm$0.15    & 0.27$\pm$0.05 & 4.59$\pm$0.16\\
07/02/08 & 2.8799 & 6.40$\pm$0.19    & 0.80$\pm$0.16 & 5.60$\pm$0.25\\
08/02/08 & 2.8841 & 6.26$\pm$0.15    & 0.81$\pm$0.16 & 5.45$\pm$0.22\\
09/02/08 & 2.8885 & 6.39$\pm$0.18    & 0.81$\pm$0.16 & 5.58$\pm$0.24\\
10/02/08 & 2.8927 & 6.71$\pm$0.16    & 0.82$\pm$0.16 & 5.89$\pm$0.23\\
\hline\hline
\end{tabular}
\end{table}

\acknowledgements{M. Vanko acknowledges support from the EU in the FP6 MC ToK project
MTKD-CT-2006-042514. M. M. Hohle acknowledges partial support from DFG in the SFB/TR-7 Gravitation
Wave Astronomy. We make use of data products from the Two Micron All Sky Survey, which is a joint
project of the University of Massachusetts and the Infrared Processing and Analysis
Center/California Institute of Technology, funded by the National Aeronautics and Space
Administration and the National Science Foundation, as well as the SIMBAD and VIZIER databases,
operated at CDS, Strasbourg, France.}

\begin{appendix}

\section{Appendix}

\begin{table}[h!]
\centering\caption{CTK astrometry of comet 17/P Holmes. The averaged uncertainty of the absolute
CTK astrometry is 0.28\,arcsec relative to the 2MASS point source catalogue.} \label{tabastro}
\end{table}

\tabletail{\hline}

\begin{supertabular}{lcc}

\hline
date                       & RA          & Dec        \\
yyyy\,\,\,mm\,\,\,dd.ddddd & hh:mm:ss.ss & dd:mm:ss.s \\
\hline
2007\,\,\,10\,\,\,\,\,\,29.05340 & 03:49:11.41 & +50:23:56.9\\
2007\,\,\,10\,\,\,\,\,\,29.06932 & 03:49:10.41 & +50:23:59.9\\
2007\,\,\,10\,\,\,\,\,\,29.07134 & 03:49:10.31 & +50:24:00.3\\
2007\,\,\,10\,\,\,\,\,\,29.07337 & 03:49:10.16 & +50:24:00.6\\
2007\,\,\,10\,\,\,\,\,\,29.07610 & 03:49:10.00 & +50:24:01.2\\
2007\,\,\,10\,\,\,\,\,\,29.07817 & 03:49:09.87 & +50:24:01.7\\
2007\,\,\,10\,\,\,\,\,\,29.08020 & 03:49:09.75 & +50:24:01.9\\
2007\,\,\,10\,\,\,\,\,\,29.08216 & 03:49:09.64 & +50:24:02.6\\
2007\,\,\,10\,\,\,\,\,\,29.08413 & 03:49:09.52 & +50:24:02.6\\
2007\,\,\,10\,\,\,\,\,\,29.08610 & 03:49:09.39 & +50:24:03.5\\
2007\,\,\,10\,\,\,\,\,\,29.10067 & 03:49:08.48 & +50:24:06.0\\
2007\,\,\,11\,\,\,\,\,\,27.79446 & 03:14:41.66 & +49:20:45.1\\
2007\,\,\,11\,\,\,\,\,\,27.79563 & 03:14:41.62 & +49:20:44.4\\
2007\,\,\,11\,\,\,\,\,\,27.79681 & 03:14:41.56 & +49:20:44.7\\
2007\,\,\,11\,\,\,\,\,\,27.79792 & 03:14:41.44 & +49:20:44.6\\
2007\,\,\,11\,\,\,\,\,\,27.79941 & 03:14:41.37 & +49:20:44.6\\
2007\,\,\,11\,\,\,\,\,\,27.80903 & 03:14:40.75 & +49:20:40.8\\
2007\,\,\,11\,\,\,\,\,\,28.74995 & 03:13:42.78 & +49:13:56.8\\
2007\,\,\,11\,\,\,\,\,\,28.75102 & 03:13:42.72 & +49:13:55.6\\
2007\,\,\,11\,\,\,\,\,\,28.75209 & 03:13:42.63 & +49:13:54.1\\
2007\,\,\,11\,\,\,\,\,\,28.75329 & 03:13:42.58 & +49:13:53.5\\
2007\,\,\,11\,\,\,\,\,\,28.75436 & 03:13:42.52 & +49:13:53.2\\
2007\,\,\,11\,\,\,\,\,\,28.75543 & 03:13:42.45 & +49:13:53.2\\
\hline
date                       & RA          & Dec        \\
yyyy\,\,\,mm\,\,\,dd.ddddd & hh:mm:ss.ss & dd:mm:ss.s \\
\hline
2007\,\,\,11\,\,\,\,\,\,28.75661 & 03:13:42.44 & +49:13:52.9\\
2007\,\,\,11\,\,\,\,\,\,28.75769 & 03:13:42.34 & +49:13:52.7\\
2007\,\,\,11\,\,\,\,\,\,28.75875 & 03:13:42.18 & +49:13:51.3\\
2007\,\,\,11\,\,\,\,\,\,28.75983 & 03:13:42.18 & +49:13:51.3\\
2007\,\,\,11\,\,\,\,\,\,28.78810 & 03:13:40.38 & +49:13:39.9\\
2007\,\,\,11\,\,\,\,\,\,28.78918 & 03:13:40.26 & +49:13:39.4\\
2007\,\,\,11\,\,\,\,\,\,28.79024 & 03:13:40.21 & +49:13:39.5\\
2007\,\,\,11\,\,\,\,\,\,28.79131 & 03:13:40.14 & +49:13:38.7\\
2007\,\,\,11\,\,\,\,\,\,28.79238 & 03:13:40.06 & +49:13:38.4\\
2007\,\,\,11\,\,\,\,\,\,28.79949 & 03:13:39.62 & +49:13:35.8\\
2007\,\,\,11\,\,\,\,\,\,28.80056 & 03:13:39.61 & +49:13:35.0\\
2007\,\,\,11\,\,\,\,\,\,28.80163 & 03:13:39.50 & +49:13:34.6\\
2007\,\,\,11\,\,\,\,\,\,28.80271 & 03:13:39.45 & +49:13:33.8\\
2007\,\,\,11\,\,\,\,\,\,28.80500 & 03:13:39.26 & +49:13:32.6\\
2007\,\,\,11\,\,\,\,\,\,28.80608 & 03:13:39.24 & +49:13:32.3\\
2007\,\,\,11\,\,\,\,\,\,28.80714 & 03:13:39.19 & +49:13:32.0\\
2007\,\,\,11\,\,\,\,\,\,28.80822 & 03:13:39.03 & +49:13:30.9\\
2007\,\,\,11\,\,\,\,\,\,28.80928 & 03:13:39.01 & +49:13:30.7\\
2007\,\,\,11\,\,\,\,\,\,28.81071 & 03:13:38.94 & +49:13:29.8\\
2007\,\,\,11\,\,\,\,\,\,28.81177 & 03:13:38.88 & +49:13:29.4\\
2007\,\,\,11\,\,\,\,\,\,28.81285 & 03:13:38.81 & +49:13:29.2\\
2007\,\,\,11\,\,\,\,\,\,28.81391 & 03:13:38.73 & +49:13:28.3\\
2007\,\,\,11\,\,\,\,\,\,28.81498 & 03:13:38.67 & +49:13:28.0\\
2007\,\,\,12\,\,\,\,\,\,01.88733 & 03:10:39.08 & +48:50:00.5\\
2007\,\,\,12\,\,\,\,\,\,01.88840 & 03:10:39.03 & +48:50:00.1\\
2007\,\,\,12\,\,\,\,\,\,01.88948 & 03:10:38.93 & +48:49:59.2\\
2007\,\,\,12\,\,\,\,\,\,01.89079 & 03:10:38.87 & +48:49:58.9\\
2007\,\,\,12\,\,\,\,\,\,01.89186 & 03:10:38.83 & +48:49:57.9\\
2008\,\,\,02\,\,\,\,\,\,07.80910 & 03:23:24.35 & +39:28:15.7\\
2008\,\,\,02\,\,\,\,\,\,07.81730 & 03:23:23.72 & +39:27:58.8\\
2008\,\,\,02\,\,\,\,\,\,08.79417 & 03:24:27.82 & +39:23:20.2\\
2008\,\,\,02\,\,\,\,\,\,09.78072 & 03:25:32.20 & +39:18:30.7\\
2008\,\,\,02\,\,\,\,\,\,09.78676 & 03:25:32.52 & +39:18:29.2\\
2008\,\,\,02\,\,\,\,\,\,09.79315 & 03:25:33.07 & +39:18:27.2\\
2008\,\,\,02\,\,\,\,\,\,10.77179 & 03:26:37.98 & +39:13:46.1\\
\end{supertabular}

\begin{figure*}[h!]
\resizebox{\hsize}{!}{\includegraphics[]{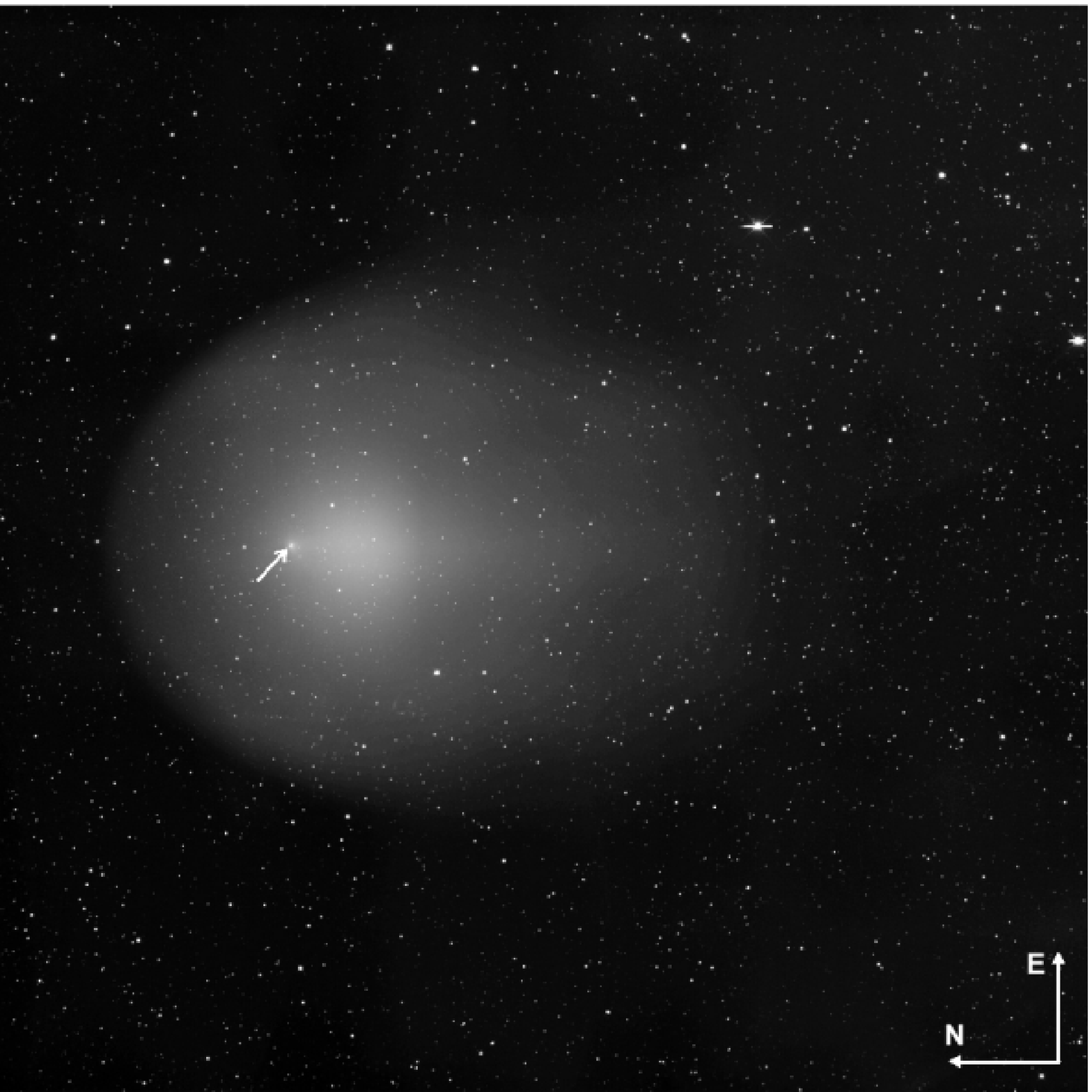}} \caption{Comet 17P/Holmes on Dec
1$^{\rm th}$ 2007. This pattern shows a V-band mosaic, composed of 9 CTK images each with an
integration time of 300\,s. The field of view covered by this mosaic is
1.57\,$^{\circ}$$\times$1.57\,$^{\circ}$, showing the huge extent of the comet's coma about one
month after the outburst in brightness. The position of the comet's central coma is indicated by a
white arrow in the CTK mosaic.} \label{holmes011207}
\end{figure*}

\end{appendix}

\end{document}